\documentstyle[aps,twocolumn,prl,amssymb]{revtex}
 \tighten
\begin{document}
\twocolumn[\hsize\textwidth\columnwidth\hsize\csname @twocolumnfalse\endcsname
\draft
\title{Correcting Quantum Errors In Higher Spin Systems}
\author{H. F. Chau\footnote{e-mail: hfchau@hkusua.hku.hk}}
\address{Department of Physics, University of Hong Kong, Pokfulam Road, Hong
Kong}
\date{\today}
\preprint{HKUPHYS-HFC-01; quant-ph/9610023}
\maketitle
\begin{abstract}
 I consider the theory of quantum error correcting code (QECC) where each
 quantum particle has more than two possible eigenstates. In this higher spin
 system, I report an explicit QECC that is related to the symmetry group
 ${\Bbb Z}_2^{\otimes (N-1)} \otimes S_N$. This QECC, which generalizes Shor's
 simple majority vote code, is able to correct errors arising from exactly one
 quantum particle. I also provide a simple encoding algorithm.
\end{abstract}
\medskip
]
\narrowtext
\footnotetext[1]{e-mail: hfchau@hkusua.hku.hk}
 Quantum computers are powerful enough to efficiently factorize composite
 numbers \cite{Shor94}. Nevertheless, quantum computers are extremely
 vulnerable to disturbance \cite{Landauer94}. Decoherence between the quantum
 computer and the environment, together with decoherence between different
 parts of a quantum computer may seriously affect the output of a computation.
\par
 By encoding the quantum state into a larger Hilbert space $H$, it is possible
 to reduce the decoherence error with the environment. By first measuring the
 wavefunction in a suitable subspace $C$ of $H$ and then by applying a unitary
 transformation to the orthogonal complement of $C$ according to the
 measurement result, it is possible to correct quantum errors due to
 decoherence with the environment \cite{Shor95}. This kind of scheme is now
 called quantum error correction code (QECC). The first QECC was discovered by
 Shor. Using the idea of simple majority vote, he encodes each quantum bit
 (qubit) by 9 qubits. His code is able to correct one qubit of error
 \cite{Shor95}. Since then, many QECCs have been discovered (see, for example,
 Refs.~\cite{Laflamme96,CS96,St96a,Steane96,Calderbank96,Gottesman96}) and
 various theories on QECC have also been developed (see, for example,
 Refs.~\cite{Steane96,Calderbank96,Gottesman96,Steane96a,Knill96a,Knill96,Bennett96}).
 In particular, the necessary and sufficient condition for a QECC is
 \cite{Knill96a,Knill96,Bennett96}
\begin{equation}
 \langle i_{\rm Encode} | A^{\dag} \! B |j_{\rm Encode} \rangle = \lambda_{A,B}
 \delta_{ij} ~, \label{E:Condition}
\end{equation}
 where $|i_{\rm Encode} \rangle$ denotes the encoded quantum state $|i\rangle$
 using the QECC, $A, B$ are the possible errors that can be handled by the
 QECC, and $\lambda_{A,B}$ is a complex constant independent of
 $|i_{\rm Encode}\rangle$ and $|j_{\rm Encode}\rangle$.
\par
 Early QECCs concentrate on the decoherence of a quantum computer with the
 environment. Individual quantum registers in a quantum computer are assumed to
 be placed far apart from each other so that decoherence between them can be
 ignored. Nonetheless, this assumption is not true in general. To understand
 why, let me first summarize the simplest possible spin-1/2 particle based
 quantum computer model below: A single spin-1/2 particle (A) is used as a
 messenger. It shuttles around other spin-1/2 particles (B) and interacts with
 them from time to time. Although decoherence between particles (B) may be
 neglected, decoherence between (A) and (B) can be serious (compare with a
 similar ``gearbox quantum computer'' proposal by DiVincenzo
 \cite{DiVincenzo95}).
\par
 Therefore, it is natural to construct QECC which corrects this kind of
 ``internal'' decoherence error between different quantum registers. This can
 be achieved by constructing QECC that may correct errors involving multiple
 spins (see, for example, Refs.~\cite{CS96,St96a,Calderbank96,Steane96a}).
 Alternatively, we may map this problem to that of correcting single quantum
 error in a system with higher spin. Suppose the messenger (A) has to interact
 with a specific spin-1/2 register (C) in (B). We may regard the combination of
 (A) and (C) as a single quantum particle with spin~3/2. If we encode this
 spin-3/2 state by an QECC and correct the quantum error immediately after the
 interaction process, decoherence between (A), (C) and the environment can be
 greatly suppressed. The advantage of this method is, in general, fewer quantum
 registers are required. The reason is simple: resources are concentrated on
 correcting errors in (A) and (C), while extra resources are needed for a
 general multiple quantum error correcting code in order to take care of the
 less frequent decoherence error within (B).
\par
 Another reason to consider QECC for higher spin system is that quantum
 registers used may consist of more than two possible states. For example, the
 two bit quantum logic gate experimentally studied by Monroe {\em et al.} uses
 extra states for preparation and measurement \cite{Monroe95}. Error
 correction may be required to prevent the quantum register from going to the
 unwanted states during the computation.
\par
 In this paper, I consider QECC for particles with spin higher than 1/2. I
 study a special kind of QECC that is related to the symmetry group
 ${\Bbb Z}_2^{\otimes (N-1)} \otimes S_N$ where $N$ is the number of states of
 each spin. An explicit example of an QECC which is able to correct one quantum
 register\footnote{Note that the state of each quantum register spans an 
 $N$-dimensional Hilbert space. When $N > 2$, it is not appropriate to call it a
 qubit because the quantum register holds more information than one qubit.} of
 error is given. My code reduces to the simple majority vote code proposed by
 Shor \cite{Shor95} when $N = 2$.
\par
 I denote the $N$ mutually orthogonal eigenstates in each quantum register by
 $|0\rangle, |1\rangle, \ldots ,|N-1\rangle$. Any quantum error involving
 exactly one quantum register can be described by an operator $E$ acting on
 that quantum register. Clearly we can represent $E$ by a non-zero $N\times N$
 complex matrix. That is to say, $E \in {\cal A} \equiv {\Bbb C}^{N\times N}
 \backslash \{ 0 \}$. Further properties of quantum error operator can be found
 elsewhere \cite{Noisy_Channel}. It is easy to check that for any $E\in
 {\cal A}$, we can find complex numbers $\alpha, \beta_i, \gamma_{mn}$ and
 $\delta_{mn}$, not all zero, such that
\begin{equation}
 E = \alpha I_N + \sum_{i = 1}^{N-1} \beta_i R_i + \sum_{m\neq n} \left(
 \gamma_{mn} P_{mn} + \delta_{mn} Q_{mn} \right) ~, \label{E:Decomposition}
\end{equation}
 where the sum in the third term runs from $m,n = 0$ to $N-1$, $I_N$ is the $N
 \times N$ identity matrix, and $R_i$, $P_{mn}$, $Q_{mn}$ are given by
\begin{mathletters}
\begin{equation}
 \left( R_i \right)_{xy} = \left\{ \begin{array}{rl} 1 & \hspace{0.2in}
 \mbox{if~} x = y \mbox{~and~} x \neq i \\ -1 & \hspace{0.2in} \mbox{if~} x =
 y = i \\ 0 & \hspace{0.2in} \mbox{otherwise} \end{array} \right. ~,
 \label{E:R_i_Def}
\end{equation}
\begin{equation}
 \left( P_{mn} \right)_{xy} = \left\{ \begin{array}{rl} 1 & \hspace{0.2in}
 \mbox{if~} x = y \mbox{~and~} x \neq m, n \\ 1 & \hspace{0.2in} \mbox{if~}
 x = m, y = n \mbox{~or~} x = n, y = m \\ 0 & \hspace{0.2in} \mbox{otherwise}
 \end{array} \right. ~, \label{E:P_mn_Def}
\end{equation}
\par\noindent
and
\begin{equation}
 \left( Q_{mn} \right)_{xy} = \left\{ \begin{array}{rl} 1 & \hspace{0.2in}
 \mbox{if~} x = y \mbox{~and~} x \neq m, n \\ 1 & \hspace{0.2in} \mbox{if~}
 x = m, y = n \\ -1 & \hspace{0.2in} \mbox{if~} x = n, y = m \\ 0 &
 \hspace{0.2in} \mbox{otherwise} \end{array} \right. ~, \label{E:Q_mn_Def}
\end{equation}
\end{mathletters}
\par\noindent
respectively. Physically, $R_i$ adds a phase shift of $\pi$ to the part of the
 state ket whenever the quantum register is in the state $|i\rangle$. The
 action of $P_{mn}$ interchanges $|m\rangle$ with $|n\rangle$ while leaving the
 other quantum states unchanged. Similarly, $Q_{mn}$ maps $|m\rangle$ to
 $|n\rangle$ and $|n\rangle$ to $-|m\rangle$ while leaving the other quantum
 states unchanged. Therefore, $R_{i}$ and $P_{mn}$ model the effect of phase
 error and spin flip, respectively. And $Q_{mn}$ models the effect of combined
 phase and spin flip error. Note that $I_N$, $R_i$, $P_{mn}$, and $Q_{mn}$ are
 Hamiltonian operators and hence, are physical observables. Besides, they form
 a linearly independent set.
\par
 From Eq.~(\ref{E:Decomposition}), it is easy to show that an QECC can handle
 one quantum register of error if and only if it can handle errors arising from
 the actions of $R_i$, $P_{mn}$ and $Q_{mn}$. Using the group theoretic method
 of QECC developed by Calderbank {\em et al.} \cite{Calderbank96}, I consider
 the finite group $G$ generated by the elements $R_i$, $P_{mn}$ and $Q_{mn}$.
 Since $P_{mn} = P_{0m} \circ P_{0n} \circ P_{0m}$, $Q_{mn} = P_{0m} \circ
 Q_{0n} \circ P_{0m}$, and $Q_{0n} = R_{n} \circ P_{0n}$, the group $G$ is
 given by
\begin{equation}
 G = \langle R_1, R_2, \ldots R_{N-1}, P_{01}, P_{02}, \ldots P_{0\,N-1}
 \rangle ~. \label{E:Group}
\end{equation}
Thus, $G$ is isomorphic to ${\Bbb Z}_2^{\otimes (N-1)} \otimes S_N$. According
 to Knill \cite{Knill96}, this choice of error bases is ``nice'' but not ``very
 nice'' in general.
\par
 Eq.~(\ref{E:Group}) implies that the ability to correct the $2(N-1)$ kinds of
 quantum errors $R_n$ and $P_{1n}$ ($n = 1,2,\ldots ,N-1$) is a necessary
 condition for correcting any quantum errors involving one quantum register.
 Here, I show that this condition is also sufficient. As shown by Gottesman
 \cite{Gottesman96}, we may paste QECC as follows: Suppose $C_1$ and $C_2$ are
 two QECCs correcting errors $E_1$ and $E_2$, respectively. Let us consider the
 situation when both errors occur in the same set of quantum registers. One can
 first encode the quantum register using code $C_1$, and then further encode
 the resultant quantum registers by the code $C_2$. The resultant quantum code
 can correct errors in the form $E_2 \circ E_1$. Thus, by pasting QECCs that
 corrects the quantum errors $R_n$ and $P_{1n}$ ($n = 1,2,\ldots ,N-1$) in a
 suitable way, one obtains a QECC for quantum errors given by the group $G$,
 and hence this code corrects quantum errors involving exactly one quantum
 register.
\par
 Since the coding scheme
\begin{equation}
 |i\rangle \longmapsto |iii\rangle \label{E:Flip_Encode}
\end{equation}
 can correct quantum errors $P_{mn}$, and the coding scheme
\begin{eqnarray}
 |1\rangle & \longmapsto & \frac{1}{\sqrt{8}} \left( |1\rangle + |i\rangle
 \right) \otimes \left( |1\rangle + |i\rangle \right) \otimes \left( |1\rangle
 + |i\rangle \right) ~, \nonumber \\ |i\rangle & \longmapsto &
 \frac{1}{\sqrt{8}} \left( |1\rangle - |i\rangle \right) \otimes \left(
 |1\rangle - |i\rangle \right) \otimes \left( |1\rangle - |i\rangle \right) ~,
 \nonumber \\ |j\rangle & \longmapsto & |jjj\rangle
\end{eqnarray}
 can correct the quantum error $R_i$. One may paste these codes together to
 obtain the required QECC that can correct errors involving one quantum
 register. Nevertheless, this construction is not practical since it involves
 too many quantum registers.
\par
 Here, I report a more economical code. Suppose $\omega_N$ is a primitive
  $N$-th root of unity, then
\begin{equation} 
 \sum_{m=0}^{N-1} \omega_N^{mk} = \left\{ \begin{array}{ll} 0 & \hspace{0.2in}
 \mbox{for~} k = 1,2,\ldots ,N-1 \\ N & \hspace{0.2in} \mbox{if~} k = N
 \end{array} \right. ~. \label{E:omega}
\end{equation}
 Consequently, state kets $|0\rangle + \omega_N^k |1\rangle + \omega_N^{2k}
 |2\rangle + \ldots + \omega_N^{(N-1)k} |N-1\rangle$ are mutually orthogonal
 to each other for $k = 0,1,\ldots ,N-1$. Besides, one can verify that the
 encoding
\begin{eqnarray}
 |m\rangle & \longmapsto & \frac{1}{N^{3/2}} \left[ \sum_{k=0}^{N-1}
 \omega_N^{km} |k\rangle \right] \otimes \left[ \sum_{k=0}^{N-1}
 \omega_N^{km} |k\rangle \right] \otimes \nonumber \\ & & ~~~\left[
 \sum_{k=0}^{N-1} \omega_N^{km} |k\rangle \right] \nonumber \\ & = &
 \frac{1}{N^{3/2}} \sum_{k,p,q = 0}^{N-1} \omega_N^{(k+p+q)m} |kpq\rangle
 \label{E:Phase_Encode}
\end{eqnarray}
 can correct phase quantum errors $R_i$ ($i=1,2,\ldots ,N-1$).
\par
 Since $R_i$ commutes with $P_{mn}$, so by pasting the two codes in
 Eqs.~(\ref{E:Flip_Encode}) and~(\ref{E:Phase_Encode}) together, we obtain
 a QECC that handles errors in $G$ (see Ref.~\cite{Gottesman96}). I explicitly
 write down this code below:
\begin{eqnarray}
 |m\rangle & \longmapsto & \frac{1}{N^{3/2}} \left[ \sum_{k=0}^{N-1}
 \omega_N^{km} |kkk\rangle \right] \otimes \left[ \sum_{k=0}^{N-1}
 \omega_N^{km} |kkk\rangle \right] \otimes \nonumber \\ & & ~~~\left[
 \sum_{k=0}^{N-1} \omega_N^{km} |kkk\rangle \right] \nonumber \\ & = &
 \frac{1}{N^{3/2}} \sum_{k,p,q=0}^{N-1} \omega_N^{(k+p+q)m} |kkkpppqqq\rangle
 \label{E:Encode}
\end{eqnarray}
 for all $m = 0,1,2,\ldots ,N-1$. Note that this code encodes each quantum
 register by nine of them, and it is able to correct any quantum errors arising
 from exactly one quantum register. When $N = 2$, it reduces to the simple
 majority code by Shor \cite{Shor95}.
\par
 The above QECC is closely related to the (multiplicative) group character
 $\chi$ of the finite additive group ${\Bbb Z}_N$. Note that $\chi : {\Bbb Z}_N
 \longrightarrow {\Bbb C}$ is a map satisfying \cite{NT}
\begin{equation}
 \chi (a + b) = \chi (a) \chi (b) \label{E:Group_Character}
\end{equation}
 for all $a,b\in {\Bbb Z}_N$. If we identify each eigenstate $|m\rangle$ with
 $m\in {\Bbb Z}_N$, then Eq.~(\ref{E:omega}) is a direct consequence of the sum
 rule \cite{NT}
\begin{equation}
 \sum_{m\in {\Bbb Z}_N} \chi (m) = \left\{ \begin{array}{ll} N & \hspace{0.2in}
 \mbox{if~} \chi \mbox{~is~the~trivial~character} \\ 0 & \hspace{0.2in}
 \mbox{otherwise} \end{array} \right. ~. \label{E:Sum_Rule}
\end{equation}
 The above sum rule ensures that the encoded states $|m_{\rm Encode}\rangle$
 given by Eq.~(\ref{E:Phase_Encode}) are mutually orthogonal.
\par
 Now, I provide a simple encoding algorithm for this code. Using a series of
 quantum binary conditional-NOT gates, we may ``copy'' the quantum state
 $|m00000000\rangle$ to $|m00m00m00\rangle$ efficiently. Then, we may apply
 quantum discrete Fourier transform similar to that used in the Shor's
 factorization algorithm \cite{Shor94,Ekert96} separately to the first, fourth,
 and the seventh quantum registers in order to produce the required encoding
 scheme. That is to say, for each $|m\rangle$ in the first, fourth, and the
 seventh quantum registers, we apply a unitary transformation, mapping it to
 the state
\begin{equation}
 |m\rangle \longmapsto \frac{1}{\sqrt{N}} \sum_{k=0}^{N-1} \omega_N^{km}
 |k\rangle ~. \label{E:DFT}
\end{equation}
Using the same idea as in the Shor's algorithm, the above transformation can
 be achieved efficiently. To obtained the required encoding, we finally
 ``copy'' the first quantum register into the second and the third, the fourth
 into the fifth and the sixth, and the seventh into the eighth and the ninth.
 The entire process can be summarized below
\begin{eqnarray}
 |m00000000\rangle & \longmapsto & |m00m00m00\rangle \nonumber \\ & \longmapsto
 & \frac{1}{N^{3/2}} \sum_{k,p,q=0}^{N-1} \omega_N^{(k+p+q)m} |k00p00q00\rangle
 \nonumber \\ & \longmapsto & \frac{1}{N^{3/2}} \sum_{k,p,q=0}^{N-1}
 \omega_N^{(k+p+q)m} |kkkpppqqq\rangle ~. \label{E:Encode_Method}
\end{eqnarray}
\par\indent
 In order to have enough room in the encoded Hilbert space for the QECC, the
 condition
\begin{equation}
 \left[ 1 + (N^2 - 1) n \right] N \leq N^n \label{E:Dimension}
\end{equation}
 must be satisfied, where $n$ is the number of quantum register. Moreover, the
 code is said to be perfect if the equality in Eq.~(\ref{E:Dimension}) holds
 \cite{Laflamme96}. Nonetheless, Eq.~(\ref{E:Encode}) is not a perfect code,
 and a more efficient QECC may exist. It will be interesting to find them out.

\end{document}